\documentclass [12pt,a4paper]{article}

\usepackage{amssymb, theorem}
\newtheorem{lemma}{Lemma}
\newtheorem{prop}{Proposition}
\newtheorem{thm}{Theorem}
\newtheorem{coro}{Corollary}
\theorembodyfont{\rm}
\newtheorem{ex}{Example}
\def\Ae{\mathcal A}
\def\A0{\mathcal A_0}
\def\Pe{\mathcal P}
\def\bbff{\bf }
\def\qed{{\hfill $\square$}}
\def\abs{< \!\! <}
\def\fel{{\textstyle {1 \over 2}}}
\newcommand{\<}{\langle}
\renewcommand{\>}{\rangle}
\renewcommand{\Im}{\hbox{Im}}
\renewcommand{\Re}{\hbox{Re}}
\def\b1{{\bf 1}}
\def\im{{\rm i}}
\def\Prob{{\rm Prob}}
\def\iM{\mathcal M}
\def\iL{\mathcal L}
\def\iA{\mathcal A}

\def\iH{\mathcal H}
\def\iK{\mathcal K}
\def\iE{\mathcal E}
\def\iN{\mathcal N}
\def\iP{\mathcal P}

\def\iH{\mathcal H}
\def\iS{\mathcal S}
\def\iX{\mathcal X}

\def\vfi{\varphi}
\def\aa{\alpha}

\def\bbbr{{\mathbb R}}
\def\bbbc{{\mathbb C}}
\def\CCR{{\rm CCR}}

\def\supp{\mbox{supp}\,}
\def\Tr{\mathrm Tr}

\def\sig{\sigma}
\def\ssum{\textstyle{\sum}}

\begin{document}
\ \vskip 1cm
\centerline{\LARGE Sufficiency in quantum statistical inference.}
\medskip
\centerline{\LARGE A survey with examples}
\bigskip
\bigskip
\bigskip
\centerline{\large Anna Jen\v cov\'a\footnote{Supported by the EU
Research Training Network Quantum Probability with Applications to
Physics, Information Theory and Biology. E-mail: jenca@mat.savba.sk}}
\centerline{Mathematical Institute of the} \centerline{Slovak
Academy of Sciences} \centerline{Stefanikova 49, Bratislava,
Slovakia}
\bigskip
\centerline{\large D\'enes Petz\footnote{Supported by the Hungarian grant
OTKA T032662. E-mail: petz@renyi.hu}}
\centerline{Alfr\'ed R\'enyi Institute of Mathematics}
\centerline{Hungarian Academy of Sciences}
\centerline{POB 127, H-1364 Budapest, Hungary}
\bigskip

\medskip\bigskip

\begin{quote}
This paper attempts to give an overview about sufficiency in the setting
of quantum statistics. The basic concepts are treated paralelly to the
the measure theoretic case. It turns out that several classical examples
and results have a non-commutative analogue. Some of the results are presented
without proof (but with exact references) and the presentation is intended
to be self-contained. The main examples discussed in the paper are related
the Weyl algebra and to the exponential family of states. The characterization
of sufficiency in terms of quantum Fisher information is a new result.
\end{quote}

\begin{quote}
\end{quote}

\begin{quote}

\end{quote}
\bigskip\bigskip
In order to motivate the concept of sufficiency, we first turn to the setting of
classical statistics. Suppose we observe an $N$-dimensional random vector
$X$, characterised by the density function  $f(x |\theta)$, where $\theta$
is a $p$-dimensional  vector of parameters and $p < N$. Assume that the
densities $f(x|\theta)$  are known and the parameter $\theta$  completely
determines the distribution of $X$.  Therefore, $\theta$ is to be estimated.
The $N$-dimensional observation $X$ carries information about the
$p$-dimensional parameter vector $\theta$. One may ask the following
question: Can we compress $x$ into a low-dimensional statistic
without any loss of information?  Does there exist some function  $t=Tx$,
where the dimension of $t$ is less than $N$, such that $t$ carries
all the useful information about $\theta$? If so, for the purpose of studying
$\theta$,  we could discard the measurements $x$ and retain  only the
low-dimensional statistic $t$.  In this case, we call $t$ a sufficient statistic.
The following example is standard and simple. Suppose a binary information
source emits a sequence of $0$'s and $1$'s, we have the independent variables
$X_1, X_2,\dots, X_N$ such that  $\Prob (X_i= 1)=\theta$. In this case the empirical
mean
$$
T(x_1,x_2,\dots, x_N)=\frac{1}{N}\sum_{i=1}^N x_i
$$
can be used to estimate the parameter $\theta$ and it is a sufficient statistic.

\section{Preliminaries}

A quantum mechanical system is described by a C*-algebra, the dynamical
variables (or observables) correspond to the self-adjoint elements and
the physical states of the system are modelled by the normalized positive
functionals of the algebra, see \cite{BEH, BR}. The evolution of the
system $\iM$ can be described in the {\bbff Heisenberg picture} in which
an observable $A \in \iM$ moves into $\alpha(A)$, where $\alpha$ is a
linear transformation. $\alpha$ is an automorphism in case of the time
evolution of a closed system but it could be the irreversible evolution
of an open system. The {\bbff Schr\"odinger picture} is dual, it gives
the transformation of the states, the state $\vfi \in \iM^*$ moves
into $\vfi \circ \alpha$. The algebra of a quantum system is typically
non-commutative but the mathematical formalism supports commutative algebras
as well. A simple {\bbff measurement} is usually modelled by a family of pairwise
orthogonal projections, or more generally, by a partition of unity,
$(E_i)_{i=1}^n$. Since all $E_i$ are supposed to be positive and
$\sum_i E_i=I$, $\beta: \bbbc^n \to \iM$, $(z_1,z_2,\dots, z_n)\mapsto
\sum_i z_i E_i$ gives a positive unital mapping from the commutative
C*-algebra $\bbbc^n$ to the non-commutative algebra $\iM$.
Every positive unital mappings occur in this way. The essential concept
in quantum information theory is the state transformation which is affine
and the dual of a positive unital mapping. All these and several other
situations justify to study of positive unital mappings between C*-algebras
from a quantum statistical viewpoint.

If the algebra $\iM$ is ``small'' and $\iN$ is ``large'', and  the mapping
$\alpha: \iM \to \iN$  sends the state $\vfi$ of the system of interest
to the state $\vfi \circ \alpha$ at our disposal, then loss of information
takes place and the problem of statistical inference is to reconstruct the
real state from partial information. In this paper we mostly consider
parametric statistical models, a parametric family $\iS:=\{\vfi_\theta:
\theta \in \Theta \}$ of states is given and on the basis of the partial
information the correct value of the parameter should be decided. If the
partial information is the outcome of a measurement, then we have statistical
inference in the very strong sense. However, there are ``more quantum''
situations, to decide between quantum states on the basis of quantum data.
The problem we discuss is not the procedure of the decision about the true 
state of the system but we want to describe the circumstances under which 
this is perfectly possible.

In this paper, C*-algebras always have a unit $I$. Given a C*-algebra $\iM$,
a state $\vfi$ of $\iM$ is a linear function $\iM \to \bbbc$ such that
$\vfi(I)=1=\|\vfi\|$. (Note that the second condition is equivalent to
the positivity of $\vfi$.) The books \cite{BEH, BR} -- among many others --
explain the basic facts about C*-algebras. The class of finite dimensional
full matrix algebras form a small and algebraically rather trivial subclass
of C*-algebras, but from the view-point of non-commutative statistics,
almost all ideas and concepts appear in this setting. A matrix algebra
$M_n(\bbbc)$ admits a canonical trace $\Tr$ and all states are described
by their densities with respect to $\Tr$. The correspondence is given
by $\vfi(A)=\Tr \rho_\vfi A\quad$ ($A \in M_n(\bbbc)$) and we can simply identify
the functional $\vfi$ by the density $\rho_\vfi$. Note that the density is
a positive (semi-definite) matrix of trace 1.

\begin{ex}
Let $\iX$ be a finite set and $\iN$ be a C*-algebra. Assume that for
each $x \in \iX$ a positive operator $E(x)\in \iN$ is given and
$\sum_x E(x)=I$. In quantum mechanics such a setting is a model for
a measurement with values in $\iX$.

The space $C(\iX)$ of function on $\iX$ is a C*-algebra and the partition
of unity $E$ induces a coarse-graining $\alpha: C(\iX)\to \iN$ given
by $\alpha(f)=\sum_x f(x)E(x)$. Therefore a coarse-graining defined
on a commutative algebra is an equivalent way to give a measurement.
(Note that the condition of 2-positivity is automatically fulfilled on a
commutative algebra.)\qed
\end{ex}

\begin{ex}
Let $\iM$ be the algebra of all bounded operators acting on a Hilbert space
$\iH$ and let $\iN$ be the infinite tensor product $\iM \otimes \iM \otimes
\dots $. (To understand the essence of the example one does not need the
very formal definition of the infinite tensor product.) If $\gamma$
denotes the right shift on $\iN$, then we can define a sequence $\alpha_n$
of coarse-grainings $\iM \to \iN$:
$$
\alpha_n(A):= \frac{1}{n}\big(A+\gamma(A)+\dots +\gamma^{n-1}(A)\big).
$$
$\alpha_n$ is the quantum analogue of the {\bbff sample mean}. \qed
\end{ex}

In this survey paper, the emphasis is put on the definitions and on the results.
The results obtained in earlier works are typically not proved but several 
examples are presented to give a better insight. Fisher information is a simple
an widely used concept in classical statistics. The relation of sufficiency
and quantum Fisher information is new and proved here in details. (However,
the concept of quantum Fisher information is rather concisely discussed.)
 
\section{Basic definitions}

In this section we recall some well-known results from classical
mathematical statistics, \cite{Strasser} is our general reference,
and the basic concepts of the quantum cases are discussed paralelly.

Let $(X_i,\Ae_i,\mu_i)$ be probability spaces ($i=1,2$).
Recall that a positive linear map $M:\ L^{\infty}(X_1,\Ae_1,\mu_1)\to
L^{\infty}(X_2,\Ae_2,\mu_2)$ is called a {\bbff Markov operator} if  it
satisfies $M1=1$ and $f_n\searrow 0$ implies $Mf_n\searrow 0$.

Let $\iM$ and $\iN$ be C*-algebras. Recall that {\bbff 2-positivity} of
$\alpha: \iM \to \iN$ means that
$$
\left[ \begin{array}{cc}
\alpha(A)&\alpha(B)\\ \alpha(C)&\alpha(D)\end{array} \right] \geq 0
\qquad \hbox{\ if }\qquad
\left[ \begin{array}{cc}
A& B\\ C& D \end{array} \right] \geq 0\,
$$
for $2\times 2$ matrices with operator entries. It is well-known that
a 2-positive unit-preserving mapping $\alpha$ satisfies the {\bbff Schwarz
inequality}
\begin{equation}\label{E:schwarz}
\alpha(A^*A)\geq \alpha(A)^*\alpha(A).
\end{equation}

A 2-positive unital mapping between C*-algebras will be called
{\bbff coarse-graining}. All Markov operators (defined above) are
coarse-grainings. For mappings defined between von Neumann algebras,
the monotone continuity is called {\bbff normality}. When $\iM$ and
$\iN$ are von Neumann algebras, a coarse-graining $\iM \to \iN$
will be always supposed to be normal. Therefore, our concept of
coarse-graining is the analogue of the Markov operator.

We mostly mean that a coarse-graining  transforms observables to observables
corresponding to the {\bbff Heisenberg picture} and in this case we assume that
it is unit preserving. The dual of such a mapping acts on states or on density
matrices and it will be called state transformation.

Let $(X,{\mathcal  A})$ be a measurable space and let ${\iP }=
\{P_{\theta}:\ \theta\in \Theta\}$ be a set of probability measures
on $(X,\Ae)$. Usually, $\iP$ is called {\bbff statistical experiment},
if it contains only two measures, then we speak about a {\bbff binary
experiment}. The aim of estimation theory is to decided about the
true value of $\theta$ on the basis of data.

A  sub-$\sigma$-algebra $\iA_0 \subset \iA$ is {\bbff sufficient } for
the family $\iP$ of measures if for all $A\in \iA$, there is an
$\iA_0$-measurable function $f_A$ such that for all $\theta$,
$$
f_A=P_{\theta}(A|\mathcal A_0)\quad P_{\theta}\mbox{-almost everywhere},
$$
that is,
\begin{equation}
P_{\theta}(A\cap A_0)=\int_{A_0}f_AdP_{\theta}
\end{equation}
for all $A_0\in \iA_0$ and for all $\theta$. It is clear from this
definition that if $\A0$ is sufficient then for all $P_{\theta}$
there is a common version of the conditional expectations
$E_{\theta}[g|\A0]$ for any measurable step function
$g$, or, more generally, for any function $g\in \cap_{\theta
\in \Theta}L^1 (X,\Ae,P_{\theta})$.

In the most important case, the family $\iP $ is {\bbff dominated}, that is
there is a $\sigma$-finite measure $\mu$ such that $P_\theta$ is absolutely
continuous with respect to $\mu$ for all $\theta$, this will be denoted
by $\iP \abs \mu$. A finite family is always dominated.

For our purposes, it is more suitable to use the following characterization of
sufficiency in terms of randomisation.

Let $\Pe_i =\{P_{i,\theta}:\theta\in \Theta\}$ be dominated families
of probability
measures on $(X_i,\Ae_i)$, such that $\Pe_i\equiv\mu_i$, $i=1,2$. We say that
$(X_2,\Ae_2,\Pe_2)$ is a {\bbff randomisation} of $(X_1,\Ae_1,\Pe_1)$, if there
exists a Markov operator
$M:\ L^{\infty}(X_2,\Ae_2,\mu_2) \to L^{\infty}(X_1,\Ae_1,\mu_1)$,
satisfying
$$
\int (Mf)dP_{\theta,1}=\int fdP_{\theta,2} \qquad (\theta\in\Theta,\
f\in L^{\infty}(X_2,\Ae_2,\Pe_2)).
$$
If also $(X_1,\Ae_1,\Pe_1)$ is a randomisation of  $(X_2,\Ae_2,\Pe_2)$, then
$(X_1,\Ae_1,\Pe_1)$ and $(X_2,\Ae_2,\Pe_2)$ are {\bbff statistically
equivalent}.

For example, let $\Pe\equiv P_0$ and let $\A0\subseteq \Ae$ be a subalgebra.
Then $(X,\A0,\Pe|_{\A0})$ is obviously a randomisation of $(X,\Ae,\Pe)$, where
the Markov operator is the inclusion
$L^{\infty}(X,\A0,P_0|_{\A0})\to L^{\infty}(X,\Ae,P_0)$. On the
other hand, if $\A0$ is sufficient, then the map
$$
f\mapsto E[f|\A0],\qquad E[f|\A0]=E_{\theta}[f|\A0],\quad P_{\theta}
\mbox{-almost everywhere},
$$
is a Markov
operator $L^{\infty}(X,\Ae,P_0)\to L^{\infty}(X,\A0,P_0|_{\A0})$ and
$$
\int E[f|\A0]dP_{\theta}|_{\A0}=\int fdP_{\theta}
\qquad (f\in L^{\infty}(X,\Ae,P_0),
\ \theta\in\Theta).
$$
We have the following characterizations of sufficient sub-$\sigma$-algebras.

\begin{prop} Let $\Pe$ be a dominated family and let $\A0\subseteq \Ae$
be a sub-$\sigma$-algebra. The following are equivalent.
\begin{enumerate}
\item[(i)] $\A0$ is sufficient for $\Pe$
\item[(ii)] There exists a measure $P_0$ such that $\Pe\equiv P_0$ and
${dP_{\theta}/}{dP_0}$ is $\A0$-measurable for all $\theta$.
\item[(iii)] $(X,\Ae,\Pe)$ and  $(X,\A0,\Pe|_{\A0})$ are
statistically equivalent
\end{enumerate}
\end{prop}

A classical {\bbff sufficient statistic} for the family $\iP$ is a
measurable mapping $T:\ (X,\mathcal A)\to (X_1,\mathcal A_1)$
such that  the sub-$\sigma$-algebra ${\mathcal A}^T$ generated by $T$  is 
sufficient for $\iP $. To any statistic $T$, we associate a Markov operator
$$
\tilde T:\ L^{\infty}(X_1,\Ae_1,P_0^T)\to L^{\infty}(X,\Ae,P_0),\quad
(\tilde Tg)(x)=g(T(x)).
$$
Obviously,  $(X_1,\Ae_1,\Pe^T)$ is a randomisation of $(X,\Ae,\Pe)$.
As in the case of subalgebras, we have

\begin{prop}\label{prop:clsufs}
The statistic $T:\ (X,\Ae)\to (X_1,\Ae_1)$ is sufficient for
 $\Pe$ if and only if
$(X,\Ae,\Pe)$ and $(X_1,\Ae_1,\Pe^T)$ are statistically equivalent.
\end{prop}

\begin{ex}\label{E:likelihood}
Let $P$ and $Q$ be measures on the $\sigma$-algebra $\iA$, that is,
$\{P,Q\}$ is a binary experiment which is dominated by $\mu:=P+Q$.
Let us define the function
$$
T: X\ni x\mapsto \frac {dP}{d\mu}(x)\in [0,1]
$$
$T$ is a minimal sufficient statistic for $\{P,Q\}$. For illustration,
we prove this statement directly.

Let $\iA_0\subseteq\iA$ be a sub-$\sigma$-algebra. For $A\in\iA$, let
us denote $f_A:=P_0(A|\iA_0)$. We show that $f_A$ is a common version
of $P(A|\iA_0)$ and $Q(A|\iA_0)$ if and only if $T$ is $\iA_0$-measurable.
Indeed, for $A_0\in\iA_0$,
$$
P(A\cap A_0)=\int_{A_0}\b1_A\,dP=\int_{A_0}\b1_AT\,d\mu=
\int_{A_0}E_\mu[\b1_AT|\iA_0]\,d\mu
$$
and similarly,
$$
Q(A\cap A_0)=\int_{A_0}E_{\mu}[\b1_A(1-T)|\iA_0]\,d\mu\,.
$$
The fact that $T$ is $\iA_0$-measurable is equivalent with
$$
\int_{A_0}E_{\mu}[\b1_AT|\iA_0]\,d\mu=\int_{A_0}f_AT\,d\mu=
\int_{A_0}f_AdP
$$
for all $A_0\in\iA_0$, and similarly for $Q$.

Let $p:=\frac {dP}{d\mu}$, $q:=\frac {dQ}{d\mu}$. Then
$$
\frac{dQ}{dP}:= \frac qp \b1_{\{p>0\}}\, .
$$
is called the {\bbff likelihood ratio} of $Q$ and $P$.

Since
$$
\frac{dQ}{dP}= \frac {1-T}{T}\b1_{\{T>0\}},
$$
the likelihood ratio and $T$ generates the same $\sigma$-algebra.
It follows that the likelihood ratio is a minimal sufficient statistic
as well. \qed
\end{ex}

\begin{prop}(Factorization criterion)\label{P:fact}
Let $\Pe \abs \mu$. The statistic
$T:\ (X,\Ae)\to(X_1,\Ae_1)$ is sufficient for $\Pe$  if and only if
there is an $\Ae_1$-measurable function $g_{\theta}$ for all $\theta$ and an
$\Ae$-measurable function $h$ such that
$$
\frac{dP_{\theta}}{d\mu}(x)=g_{\theta}(T(x))h(x)
\quad P_{\theta} \mbox{-almost everywhere.}
$$
\end{prop}

\begin{ex}\label{Ex:normal}
Let $X_1,X_2,\dots,X_N$ be independent random variables with normal distribution
$N(m, \sigma)$.  It is well-known that the empirical mean
$$
T(x_1,x_2,\dots, x_N)=\frac{1}{N}\sum_{i=1}^N x_i
$$
is a sufficient statistic for the parameter $m$, when $\sigma$ is fixed.

The joint distribution is
\begin{eqnarray*}
&&\prod_{i=1}^N C \exp \left(-\frac{(x_i-m)^2}{2\sigma^2}\right)\\&&=
C^N
\exp\left(-\frac{m}{\sigma^2}\sum_{i=1}^N x_i-\frac{n m^2}{2\sigma^2}\right)
\exp \left(-\frac{\sum_{i=1}^N x_i^2}{2\sigma^2}\right).
\end{eqnarray*}
and we observe the factorization:
$$
f(x,m)=g(T(x),m)h(x)
$$
According to  Proposition \ref{P:fact}, this is enough for the sufficiency. \qed
\end{ex}

Next we formulate the non-commutative setting. Let $\iM$ be a von Neumann
algebra and $\iM_0$ be its von Neumann subalgebra. Assume that a family
$\iS:=\{ \vfi_\theta: \theta \in \Theta\}$ of normal states are given.
$(\iM,\iS)$ is called {\bbff statistical experiment}. The subalgebra $\iM_0
\subset\iM$ is {\bbff sufficient} for  $(\iM,\iS)$ if for every $a \in \iM$,
there is $\aa(a) \in \iM_0$ such that
\begin{equation}
\vfi_\theta(a)=\vfi_\theta(\aa(a))  \qquad (\theta \in \Theta)
\end{equation}
and the correspondence $a \mapsto \alpha(a)$ is a coarse-graining.
(Note that a positive mapping is automatically completely
positive if it is defined on a commutative algebra.)

We will now define sufficient coarse-grainings.
Let $\iN$, $\iM$ be C*-algebras and let $\sigma:\ \iN\to \iM$ be a
coarse-graining.  By Proposition \ref{prop:clsufs}, the classical definition
of sufficiency can be generalised in the following way: we say that $\sigma$
is sufficient for the statistical
experiment $(\iM,\vfi_\theta)$ if there exists  a
coarse-graining $\beta:\iM \to \iN$ such that
$\vfi_\theta \circ \sig \circ \beta = \vfi_\theta$ for every $\theta$.

The next example is the analogue of Example \ref{Ex:normal} on the algebra of the
canonical commutation relation. Note that the bilinear form $\alpha$ plays the
role of the variance (while $\sigma$ denotes a simplectic form).

\begin{ex}\label{ex:ccr}
Let $\sigma$ be a non-degenerate symplectic form on a linear space ${\cal H}$.
Typically, ${\cal H}$ is a complex Hilbert space  and  $\sigma(f,g) =
\hbox{Im}\<f,g\>$.  The {\bbff Weyl algebra} $\CCR({\cal H})$ is
generated by unitaries  $\{W(f):\ f \in {\cal H} \}$
satisfying the Weyl form of  the {\bbff canonical commutation relation}:
$$
W(f)W(g) = e^{\im\sigma(f,g)}W(f+g)\qquad(f,g\in \iH),
$$
see the monographs \cite{BR, PDccr} about the details.
Since the linear hull of the unitaries $W(f)$ is dense in $\CCR({\cal
H}$), any state is determined uniquely by its values taken on the Weyl
unitaries.   The most important states of the Weyl algebra are the Gaussian
(or quasifree) states which are given as
$$
\varphi_{m,\alpha}(W(f)) = \exp \left(m(f) \im  - \frac{1}{2}\alpha(f,f)\right)
\qquad(f\in \iH),
$$
where $m$ is a linear functional and $\alpha$ is a bilinear functional
on $\iH$ and $\iH\times \iH$, respectively. Note that $\alpha$
should satisfy the constrain
\begin{equation}\label{E:sigma}
\sigma(f,g)^2\leq \alpha(f,f)\alpha(g,g),
\end{equation}
see Thm. 3.4 and its proof in \cite{PDccr}.

It is well-known that
$$
\CCR(\iK_1) \otimes \CCR(\iK_2)\otimes \dots \otimes
\CCR(\iK_n)
$$
may be regarded as
$$
\CCR(\iK_1\oplus \iK_2\oplus \dots \oplus \iK_n)
$$
for any Hilbert spaces $\iK_1, \iK_2, \dots \iK_n$. Now we suppose that all these
spaces coincide with  $\iH$ and we write  $\iH_n$ for  $\iH\oplus \iH\oplus \dots
\oplus \iH$.  The bilinear forms $\alpha_n$ and $\sigma_n$ defined on $\iH_n$
are induced by $\alpha$ and $\sigma$.

There exists a completely positive (so-called quasifree) mapping
$$
T:\CCR(\iH)\to \CCR(\iH_n)
$$
such that
$$
T(W(f))=W\left(\frac{1}{\sqrt{n}}(f \oplus f \oplus \dots \oplus f)\right)
$$
(p. 73 in \cite{PDccr}). We claim that $T$ is sufficient for the family
$$
\{\psi_{m,\alpha}:= \varphi_{m,\alpha}^{(1)}\otimes \varphi_{m,\alpha}^{(2)} \otimes
\dots \otimes \varphi_{m,\alpha}^{(n)}: m\}
$$
of states on $\CCR(\iH_n)$, when $\alpha$ is fixed.

Consider the quasi-free mapping $S_\alpha:\iA^{(n)} \to \CCR(\iH)$
given as
$$
S_\alpha(W(f_1\oplus f_2 \oplus \dots \oplus f_n))=
W\left(\frac{1}{\sqrt{n}}\sum_i f_i\right)
\exp \left(\frac{1}{2n}{\alpha(\ssum_i f_i,\sum_i f_i)}
-\frac{1}{2}\ssum_i \alpha(f_i,f_i)\right).
$$

Then
\begin{eqnarray*}
(T\circ S_\alpha)(W(f_1\oplus f_2 \oplus \dots \oplus f_n))&=&
W\left(\frac{1}{n}\sum_i f_i \oplus \dots \oplus \frac{1}{n}\sum_i f_i\right)
\cr && \times \exp \left(\frac{1}{2n}{\alpha(\ssum_i f_i,\sum_if_i)}
-\frac{1}{2}\ssum_i \alpha(f_i,f_i)\right)
\end{eqnarray*}
and
$$
\psi_{m,\sigma}(T\circ S_\alpha)=\psi_{m,\sigma}
$$
holds for every $m$. We will show that $S_\alpha$ is completely positive.

We can write $S_\alpha:\iA^{(n)} \to \CCR(\iH)$ as
\begin{equation}
S_\alpha(W(f^n))=W(A_nf^n)F(f^n),
\end{equation}
where $f^n=f_1\oplus\dots\oplus f_n\in \iH_n$,
$$
F(f^n)=\exp \left(\frac{1}{2n}
{\alpha(\ssum_i f_i,\ssum_i f_i)}-\frac{1}{2}\ssum_i \alpha(f_i,f_i)\right)
$$
and $A_n:\iH_n \to \iH$ is the linear map  $f_1\oplus\dots\oplus f_n\mapsto
\sum_if_i$.
By Thm. 8.1 in \cite{PDccr}, $S_\alpha$ is completely positive
if and only if the kernel
\begin{equation}\label{eq:pd}
(f^n,g^n)\mapsto F(g^n-f^n)\exp \im\Big(\sigma_n(g^n,f^n)-\sigma(A_ng^n,A_nf^n)\Big)
\end{equation}
is positive definite.

It is easy to see that $A_n^*:
f\mapsto \frac{1}{\sqrt{n}}(f \oplus f \oplus \dots \oplus f)$ and
$$
\alpha_n(f^n, (I-A_n^*A_n)g^n)=\alpha_n((I-A_n^*A_n)f^n,g^n)=
\sum_i\alpha(f_i,g_i)-\frac1{n}\alpha({\ssum_if_i,\sum_ig_i)}.
$$
Since $A_n$ is a contraction, $I-A_n^*A_n$ is positive. Setting
$B_n=(I-A_n^*A_n)^{1/2}$, we have
$$
F(f^n)=\exp \Big(-\frac 12\alpha_n\left( B_n f^n, B_n f^n\right)\Big).
$$
and
$$
\sigma_n(g^n,f^n)-\sigma(A_ng^n,A_nf^n)=
-\sigma_n(B_n f^n,B_n g^n).
$$
The kernel (\ref{eq:pd}) has the form
$$
\exp \Big(-\frac 12\alpha_n\left( B_n (g^n-f^n), B_n (g^n-f^n)+
\im \sigma_n(B_n f^n,B_n g^n)
\right)\Big).
$$
The positive definiteness follows from that of the exponent
which is so due to
$$
\sigma_n^2(f^n,g^n)\leq\alpha_n(f^n,f^n)\alpha_n(g^n,g^n).
$$ \qed
\end{ex}

\section{Sufficient subalgebras and coarse-grainings}

In the study of sufficient subalgebras monotone quasi-entropy
quantities play an important role. The {\bbff relative $\alpha$-entropies}
are examples of those \cite{petz1986b, OP}.

Let $\varphi$ and $\omega$ be normal states of a von Neumann algebra and
let $\xi_\varphi$ and $\xi_\omega$ be the representing vectors of these states
from the natural positive cone (see below). Let
$$
f_\alpha(t)={1 \over \aa(1-\aa)}\big(1-t^{\aa}\big).
$$
It is well-known that this function is operator monotone decreasing for
$\alpha \in (-1,1)$.
The relative $\alpha$-entropy
\begin{equation}\label{E:aaent}
S_\aa(\varphi|| \omega) = \<\xi_\varphi, f_\aa( \Delta)\xi_\varphi \>
\end{equation}
is a particular quasi-entropy corresponding to the function $f_\alpha$,
$\Delta$ is the relative modular operator $\Delta(\omega/\varphi)$.
When $\rho_1$ and $\rho_2$ are statistical operators, this formula
can be written as
\begin{equation}\label{E:aaent2}
S_\aa(\rho_1||\rho_2 )= {1 \over \aa(1-\aa)}
\Tr (I-\rho_2^{\aa}\rho_1^{-\aa})\rho_1\,.
\end{equation}
(For details, see Chap. 7 in \cite {OP}).

The relative $\alpha$-entropy is monotone under coarse-graining:
$$
S_\aa(\rho_1\|\rho_2) \geq S_\aa(\iE (\rho_1)\|\iE (\rho_2))\,.
$$
If follows also from the general properties of quasi-entropies that
$S_\aa(\rho_1\|\rho_2)$ is jointly convex and positive. The {\bbff transition
probability}
$$
P_A(\varphi, \omega)=\langle \xi_\varphi,\xi_\omega \rangle.
$$
corresponds to $\alpha=1/2$ (up to additive and multiplicative constans).

The next theorem is essentially Thm 9.5 from \cite{OP}.

\begin{thm}\label{thm:1}
let $\iM_0 \subset \iM$ be von Neumann algebras and let $(\iM,
\{\vfi_\theta:\theta \in \Theta\} )$ be a statistical experiment.
Assume that there are states $\vfi_n \in \iS:=\{\vfi_\theta:\theta
\in \Theta\}$ such that
$$
\omega := \sum_{n=1}^{\infty} \lambda_n \vfi_n
$$
is a faithful normal state for some constants $\lambda_n >0$.
Then the following conditions are equivalent.

\begin{enumerate}
\item[(i)]  $\iM_0$ is sufficient for $(\iM,\vfi_\theta )$.
\item[(ii)] $S_\alpha(\vfi_\theta ,\omega )=
S_\alpha(\vfi_\theta|\iM_0 ,\omega|\iM_0 )$
for all $\theta$ and for some $0 < |\aa|<1$.
\item[(iii)]  $[D\vfi_\theta , D\omega ]_t=[D(\vfi_\theta|\iM_0), D(\omega
|\iM_0)]_t\,$ for every real $t$ and for every $\theta$.
\item[(iv)] $[D\vfi_\theta,D\omega]_t\in \iM_0$ for all real $t$ and every
$\theta$.
\item[(v)]  The generalised conditional expectation
$E_\omega :\iM \to \iM_0$ leaves all the states $\vfi_\theta$ invariant.
\end{enumerate}
\end{thm}

Since $\omega$ is assumed to be
faithful and normal, it is convenient to consider a representation of $\iM$
on a Hilbert space $\iH$ such that $\omega$ is induced by a cyclic and
separating vector $\Omega$. Given a  normal state $\psi$ the quadratic
form $a\Omega \mapsto\psi (aa^*)$ ($a\in \iM$) determines the relative
modular operator $\Delta(\psi/\omega)$ as
$$
\psi (aa^*)= \|\Delta(\psi/\omega)a\Omega\|^2 \qquad (a \in \iM).
$$
The vector $\Delta(\psi/\omega)^{1/2}\Omega$ is the representative of $\psi$
from the so-called natural positive cone (which is actually the set of all
such vectors). The {\bbff Connes' cocycle}
$$
[D\psi, D\omega ]_t=\Delta(\psi/\omega)^{\im t}\Delta(\omega/\omega)^{-\im t}
$$
is a one-parameter family of contractions in $\iM$, unitaries when $\psi$ is
faithful. The modular group of $\omega$ is a group of automorphisms defined as
$$
\sigma_t(a)=\Delta(\omega/\omega)^{\im t}a\Delta(\omega/\omega)^{-\im t}
\qquad (t \in \bbbr).
$$
The Connes' cocycle is the quantum analogue of the Radon-Nikodym derivative of
measures.

The generalised conditional expectation $E_\omega: \iM\to \iM_0$ is
defined as
$$
E_\omega(a)\Omega= J_0PJa\Omega
$$
where $J$ is the modular conjugation on the Hilbert space $\iH$, $J_0$ is that
on the closure $\iH_0$ of $\iM_0\Omega$ and $P:\iH \to \iH_0$ is the orthogonal
projection \cite{AC} . There are several equivalent conditions which
guarantee that $E_\omega$ is a conditional expectation, for example,
$\sigma_t(\iM_0)\subset \iM_0$, ({\bbff Takesaki's theorem}, \cite{OP}).

More generally, let $\iM_1$ and $\iM_2$ be von Neumann algebras and let
$\sigma:\iM_1 \to \iM_2$ be a coarse-graining.
Suppose that a normal state $\vfi_2$ is given and $\vfi_1:=\vfi_2\circ \sigma$
is normal as well. Let $\Phi_i$ be the representing vectors in given natural
positive cones and $J_i$ be the modular conjugations ($i=1,2$).

{F}rom the modular theory we know that
$$
p_i:= \overline{J_i\iM_i\Phi_i}
$$
is the support projection of $\vfi_i$ (i=1,2).

The dual $\sigma_{\vfi_2}^*: p_2\iM_2 p_2 \to  p_1\iM_1 p_1$ of $\sigma$ is
is characterised by the property
\begin{equation}
\langle a_1\Phi_1, J_1\sigma_{\vfi_2}(a_2)\Phi_1\rangle =
\langle \sigma(a_1)\Phi_2, J_2 a_2\Phi_2 \rangle \quad (a_i\in \iM_i, i=1,2)
\end{equation}
(see Prop. 8.3 in \cite{OP}).

\begin{ex}\label{ex:tensor}
Let $\iM$ be a matrix algebra with a family of states $\{\vfi_\theta: \theta \in
\Theta\}$ and let  $\iM^{n \otimes}:=\iM \otimes \dots \otimes \iM$ and
$\vfi_\theta^{n \otimes}:=\vfi_\theta \otimes \dots \otimes \vfi_\theta$
be $n$-fold products. Each permutation of the tensor factors induces
an automorphism of  $\iM^{n \otimes n}$
and let $\iN$ be the fixed point subalgebra of these automorphisms. Then
$\iN$ is sufficient for the family $\{\vfi_\theta^{n \otimes}: \theta \in
\Theta\}$. Indeed, the Cones' cocycle of any two of these states is
a homogeneous tensor product, therefore they are in the fixed point algebra
$\iN$. \qed
\end{ex}

Let us return to the Weyl algebra.

\begin{ex} 
Let $\iH$ be a real Hilbert space with inner product $\alpha(f,g)$ ($f,g
\in\iH$) and let $\sigma$ be a non-degenerate symplectic form on $\iH$.
Assume that (\ref{E:sigma}) holds. Then there exists an invertible 
contraction $D$ on $\iH$, such that
$$
\sigma(f,g)=\alpha(Df,g)\qquad (f,g\in\iH).
$$
Let $D=J|D|$ be the polar decomposition, then $JD=DJ$, $J^2=-I$. The unitary
$J$ defines  a complex structure on $\iH$. We introduce a complex inner 
product by
$$
\<f,g\>:=\sigma(f,Jg)+\im \sigma(f,g),
$$
then
$$
\sigma(f,g)=\Im \<f,g\> \quad \hbox{and} \quad 
\alpha(f,g)=\Re \<|D|^{-1}f,g\>\,.
$$

For each linear form $m$ on $\iH$, there is an element $g_m\in \iH$, such
that
$$
m(f)=2\sigma(g_m,f)\qquad (f\in \iH).
$$
Let $\vfi_m$ be the quasifree state on $CCR(\iH,\sigma)$ given by
$$
\vfi_m(W(f))=\exp\left( \im m(f)-\frac12 \alpha(f,f)\right)\,.
$$
Then 
$$
\vfi_m(W(f))=\vfi_0(W(g_m)W(f)W(-g_m)) \qquad  (f\in\iH)\,.
$$

Let $H$ be a subset of $\iH$. The family of states $\iS_H=\{\vfi_m\,: \, g_m\in H\}$, 
is the quantum counterpart of the classical Gaussian shift on $\iH$.

Let us now suppose that $\|D\|<1$, then there is an  operator
$L\geq \varepsilon I$  for some $\varepsilon>0$, such that
$|D|^{-1}=\hbox{coth} L$. It was proved in \cite{PDccr} that the
state $\vfi_0$ satisfies the KMS condition with respect to the automorphism group
$$
\sigma_t(W(f))=W(V_tf)\qquad (t\in\bbbr ,\ f\in\iH)\,,
$$
where $V_t=\exp (-2\im tL )$. Therefore, $\sigma_t$ is the modular group of
$\varphi_0$. It is not difficult to prove that
$$
u_t^g=\exp (\im \sigma(V_tg,g))W(V_tg-g)\qquad (g\in \iH,\ t\in\bbbr)
$$
is the Connes' cocycle $[D\vfi_m,D\vfi_0]_t$. It follows that the  algebra 
$CCR(\iK,\sigma|\iK)$ is minimal sufficient for $\iS_\iK$ when $\iK$ is the 
subspace generated by  $\{V_t(g)-g\,:\, g\in K\}$. In particular, we see 
that if $K=\iH$, then there is no non-trivial sufficient subalgebra for the 
Gaussian shift.

Let us now recall  the situation in Example \ref{ex:ccr}. There, we studied 
the algebra $CCR(\iH_n,\sigma_n)$ with the family of states $\iS_\iL$, where
$\iL=\{g\oplus \dots \oplus g\, : \, g\in \iH\}$. It follows from our analysis
that the minimal sufficient subalgebra is $CCR(\iL,\sigma_n)$. \qed
\end{ex}

We will now define sufficient coarse-grainings.
Let $\iN$, $\iM$ be C*-algebras and let $\sigma:\ \iN\to \iM$ be a
coarse-graining.  By Proposition \ref{prop:clsufs}, the classical definition
of sufficiency can be generalised in the following way: we say that $\sigma$
is sufficient for the statistical
experiment $(\iM,\vfi_\theta)$ if there exists  a
coarse-graining $\beta:\iM \to \iN$ such that
$\vfi_\theta \circ \sig \circ \beta = \vfi_\theta$ for every $\theta$.

Let us recall the following well-known property of coarse-grainings, see
9.2 in \cite{Str}.

\begin{lemma}\label{lemma:Nsig}
Let $\iM$ and $\iN$ be C*-algebras and let $\sigma:\iN \to \iM$ be a
coarse-graining. Then
\begin{equation}
\iN_\sigma:=\{a \in \iN: \sig(a^*a)=\sig(a)\sig(a)^*\mbox{\ and\ }
\sig(aa^*)=\sig(a)^*\sig(a)\}
\end{equation}
is a subalgebra of $\iN$ and
\begin{equation}
\sig(ab)=\sig(a)\sig(b) \quad \mbox{and}\quad \sig(ba)=\sig(b)\sig(a)
\end{equation}
holds for all $a \in \iN_\sigma$ and $b \in \iN$.
\end{lemma}

We call the subalgebra $\iN_\sigma$ the {\bbff multiplicative domain}
of $\sigma$.

Now let $\iN$ and $\iM$ be von Neumann algebras and let $\omega$ be a faithful
normal state on $\iM$ such that $\omega\circ\sig$ is also faithful. Let
$$
\iN_1=\{ a\in \iN,\ \sigma_\omega^*\circ\sig(a)=a\}
$$
It was proved in \cite{petz1988} that $\iN_1$ is a subalgebra of $\iN_\sig$,
moreover, $a\in\iN_1$ if and only if $\sig(a^*a)=\sig(a)^*\sig(a)$ and
$\sig(\sig_t^{\omega\circ\sig}(a))=\sig_t^{\omega}(\sig(a))$. The restriction
of $\sig$ to $\iN_1$ is an isomorphism onto
$$
\iM_1=\{b\in\iM,\ \sig\circ\sig_\omega^*(b)=b\}
$$

The following Theorem was proved in \cite{petz1988} in the case when
$\vfi_\theta$ are  faithful states. See \cite{JP} concerning the general
case.

\begin{thm}\label{thm:3}
Let $\iM$ and $\iN$ be von Neumann algebras and let $\sig:\iN \to \iM$
be a coarse-graining. Suppose that $(\iM,\vfi_\theta )$ is a statistical
experiment dominated by a state $\omega$ such that both $\omega$ and
$\omega\circ\sig$ are faithful and normal.
Then following properties are equivalent:
\begin{enumerate}
\item[(i)] $\sig(\iN_\sig)$ is a sufficient subalgebra
for $(\iM,\vfi_\theta )$.
\item[(ii)]  $\sig$ is a sufficient coarse-graining for $(\iM,\vfi_\theta)$.
\item[(iii)] $S_\alpha(\vfi_\theta ||\omega )=S_\alpha(\vfi_\theta|\iM_0\, ||\,\omega|
\iM_0 )$ for all $\theta$ and for some $0 < |\aa|<1$.
\item[(iv)] $\sig([D\vfi_\theta\circ\sig,D\omega\circ\sig]_t)=
[D\vfi_\theta,D\omega]_t$
\item[(v)]  $\iM_1$
is a sufficient subalgebra for $(\iM,\vfi_\theta)$.
\item[(vi)] $\vfi_\theta\circ\sig\circ\sig_\omega^*=\vfi_\theta$.
\end{enumerate}
\end{thm}

The previous theorem applies to a measurement which is essentially
a positive mapping $\iN \to \iM$ from a commutative algebra. The concept
of sufficient measurement appeared also in \cite{BN-G}. For a non-commuting
family of states, there is no sufficient measurement.

We also have the following characterization of sufficient coarse-grainings in
terms of relative entropy, see \cite{petz1986}

\begin{prop}\label{prop:relent}
Under the conditions of Theorem \ref{thm:3},
suppose that $S(\vfi_\theta||\omega)$ is finite for all $\theta$. Then $\sig$ is a
sufficient coarse-graining if and only if
$$
S(\vfi_\theta ||\omega)=S(\vfi_\theta\circ\sig||\omega\circ\sig)
$$
\end{prop}

The equality in inequalities for entropy quantities was studied also in
\cite{petz03,MBR1}. For density matrices, it was shown that the equality in
Proposition \ref{prop:relent} is
equivalent to
\begin{equation}\label{E:matsuff}
\sig(\log \sig^*(D_{\theta})-\log \sig^*(D_{\omega_0}))=
\log D_\theta-\log D_\omega,
\end{equation}
where $\sig^*$ is the dual mapping of $\sig$ on density matrices.

Let us  now show how  Theorems \ref{thm:1} and \ref{thm:3} can be
applied  if the dominating state  $\omega$ is not faithful.
Suppose that  $p=\supp \omega$,  $q=\supp \omega\circ\sig$.
We  define the map $\alpha: q\iN q\to p\iM p$ by $\alpha(a)=p\sig(a)p$.
Then $\alpha$ is a coarse-graining such that $\alpha^*_\omega=\sig^*_\omega$
and $\vfi_\theta\circ \sig(a)=\vfi_\theta\circ\alpha(qaq)$ for all $\theta$.
We check that $\alpha$ is sufficient for
$(p\iM p,\vfi_\theta|_{p\iM p})$ if and only if $\sigma$ is sufficient for
$(\iM,\vfi_\theta)$. Indeed, let $\tilde\beta:p\iM p\to q\iN q$ be a
coarse-graining such that
$\vfi_\theta|_{p\iM p}\circ\alpha\circ\tilde\beta=
\vfi_\theta|_{p\iM p}$ and
let $\beta:\iM\to \iN$ be defined by
$$
\beta(a)=\tilde\beta(pap)+\omega(a)(1-q)
$$
Then $\beta$ is a coarse-graining and
$$
\vfi_\theta\circ\sig\circ\beta(a)=\vfi_\theta\circ\sig(q\beta(a)q))=
\vfi_\theta\circ\alpha\circ\tilde\beta(pap)=\vfi_\theta(pap)=\vfi_\theta(a)
$$
The converse is proved similarly,  taking $\tilde\beta(a)=q\beta(a)q$ for
$a\in p\iM p$.

\section{ Exponential families and Fisher information}

Let $\iM$ be a von Neumann algebra and $\omega$ be a normal state.
For $a \in \iM^{sa}$ define the (perturbed) state $[\omega^a]$ as 
the minimizer of the functional
\begin{equation}\label{E:min}
\psi \mapsto S(\psi|| \omega)-\psi(a)
\end{equation}
defined on normal states of $\iM$.

We define the {\bbff quantum exponential family} as
\begin{equation} \label{E:qqef}
\iS=\{\varphi_\theta:=[\omega^{\sum_i \theta_i a_i}]\, :\, \theta\in\Theta\},
\end{equation}
where $a_1,a_2,\dots,a_n$ are self-adjoint operators from $\iM$ and
$\Theta\subseteq \bbbr^n$ is the parameter space.
Let  $\iM$ be finite dimensional, and assume that  the density of
$\omega$ is written in the form $e^H$, $H=H^*\in\iM$.
Then the density of  $\vfi_\theta$ is nothing else but
\begin{equation}\label{E:qef}
\rho_\theta=\frac{\exp \left(H+ \sum_i\theta_ia_i\right)}
{\Tr\exp \left(H+ \sum_i\theta_ia_i\right) },
\end{equation}
which is a direct analogue of the classical exponential family.

Returning to the general case, note that
the support of the states $\vfi_\theta$ is $\supp \omega$.
For more details about perturbation of states, see Chap. 12 of \cite{OP},
here we recall the analogue of (\ref{E:qef}) in the general case. We 
assume that the von Neumann algebra is in a standard form and the 
representative of $\omega$ is the vector $\Omega$ from the positive cone
of the Hilbert space. Let $\Delta_\omega\equiv \Delta(\omega/\omega)$ 
be the modular operator of $\omega$, then $\varphi_\theta$ of (\ref{E:qef}) 
is the vector state induced by the unit vector
\begin{equation}
\Phi_\theta:= \frac{\exp \fel \Big(\log \Delta_\omega+ \sum_i \theta_i
a_i \Big) \Omega}{\Big\|\exp \fel\Big(\log \Delta_\omega+ \sum_i \theta_i
a_i\Big) \Omega \Big\|}\,.
\end{equation}
(This formula holds in the strict sense if $\omega$ is faithful, since
$\Delta_\omega$ is invertible in this case. For non-faithful $\omega$
the formula is modified by the support projection.) 

In the next theorem $\sigma^\omega_t$ denotes the modular automorphism
group of $\omega$, $\sigma^\omega_t(a)=\Delta_\omega^{\im t}a\Delta_
\omega^{-\im t}$.

\begin{thm} {\bf \cite{petz1986}}\label{T:4}
Let $\iM$ be a von Neumann algebra with a faithful normal state $\omega$
and $\iM_0$ be a subalgebra. For $a\in \iA^{sa}$ the
following conditions are equivalent.
\begin{enumerate}
\item[(i)] $[D[\omega^a],D\omega]_t\in \iM_0$ for all $t\in\mathbb{R}$.
\item[(ii)] $\sigma^\omega_t(a)\in \iM_0$ for all $t \in \bbbr$.
\item[(iii)] For the generalised conditional expectation $E_\omega:\iM\to
\iM_0$, $E_\omega(a)=a$ holds.
\end{enumerate}\qed
\end{thm}

\begin{coro} Let $\iS$ be the exponential family (\ref{E:qqef}) and let
$\iM_0\subseteq \iM$ be a subalgebra. Then the following are equivalent.
\begin{enumerate}
\item[(i)] $\iM_0$ is sufficient for $(\iM,\iS)$.
\item[(ii)] $\sigma^\omega_t(a_i)\in\iM_0$ for all $t\in\mathbb R$
and $1\leq i\leq n$.
\item[(iii)] $\iM_0$ is sufficient for $(\iM,\{[\omega^{a_1}],\dots,[\omega^{a_n}]\})$.
\end{enumerate}\qed
\end{coro}

Let us  denote by $c(\omega,a)$ the minimum in (\ref{E:min}), that is,
$c(\omega,a)=S([\omega^a]\,||\,\vfi)-[\omega^a](a)$. Then the function 
$a\mapsto c(\omega,a)$ is analytic and concave. 
We recall that for $a,h\in \iM^{sa}$,
$$
\frac d{dt}c(\omega,a+th)\Big|_{t=0}=-[\omega^a](h)
$$
Let us define for $a,h,k\in \iM^{sa}$
$$
\gamma_{\omega}(h,k)=-\frac{\partial^2}{\partial s\partial
t}c(\omega,th+sk)\Big|_{s=t=0}=-\frac d{dt}[\omega^{a+th}](k)\Big|_{t=0}
$$
Then $\gamma_\omega$ is a positive bilinear form on $\iM^{sa}$. It has an important monotonicity property:
If $\alpha:\iN\to\iM$ is a faithful coarse-graining, then we have for
any faithful state $\omega$ on $\iM$ and a self-adjoint element $a\in\iN$ that
$$
\gamma_\omega(\alpha(a),\alpha(a))\leq \gamma_{\omega\circ\alpha}(a,a)
$$
Note also  that for $h,k\in \iM^{sa}$ and $\lambda_1,\lambda_2\in\mathbb R$,	
$$
\gamma_\omega(h+\lambda_1,k+\lambda_2)=\gamma_\omega(h,k)
$$
and $\gamma_\omega(h,h)=0$ implies $h=\lambda\in\mathbb R$.

Let now $\iS=\{\vfi_\theta\,:\,\theta\in\Theta\}$ be a family of 
normal states on $\iM$ and  suppose that the parameter space is an 
open subset $\Theta\subset \mathbb R^k$. Further, we suppose that 
there exists a faithful normal state $\omega$ on $\iM$, such that 
there are some constants $\lambda,\mu >0$ satisfying 
\begin{equation}\label{E:family}
\lambda \omega\leq \vfi_\theta\leq \mu \omega
\end{equation}
holds for every $theta$. If this condition holds, it remains true if 
we take any element in $\iS$ in place of $\omega$, we may therefore 
suppose that $\omega\in\iS$.

Condition (\ref{E:family}) implies that
for each $\theta\in\Theta$, there is some  $a(\theta)\in \iM^{sa}$,
such that $\vfi_\theta=[\omega^{a(\theta)}]$. We will further assume that the function $\theta\mapsto a(\theta)$ is 
continuously differentiable and  denote by 
$\partial_i$ the partial derivative with respect to $\theta_i$.  

If $\alpha:\iN\to \iM$ be a coarse-graining, then for $\theta\in\Theta$,
we have
$$
\lambda\,\omega\circ\alpha\leq \vfi_\theta \circ\alpha\leq \mu
\omega\circ\alpha \,,
$$
so that the induced family again satisfies condition (\ref{E:family})
and there are self-adjoint elements $b(\theta)\in\iN$, such that
$\vfi_\theta\circ\alpha=[\omega\circ\alpha^{b(\theta)}]$.

We have the following characterization of sufficient coarse-grainings
under the above conditions

\begin{thm}\label{thm:exsuff} 
Let $\alpha:\iN\to\iM$ be a faithful coarse-graining and let $\iS$ be as above.
Then $\alpha$ is  sufficient for $(\iM,\iS)$ if and only if for each $\theta$
there is some $b(\theta)\in \iN^{sa}$, such that 
\begin{equation}\label{E:exsuff}
\vfi_\theta=[\omega^{\alpha(b(\theta))}]\quad \mbox{ and }\quad
\vfi_\theta\circ\alpha=
[\omega\circ\alpha^{b(\theta)}]. 
\end{equation}
\end{thm}

{\it Proof.} Let $\omega=\vfi_{\theta_0}\in\iS$ and let
$\vfi_\theta=[\omega^{a(\theta)}]$. 
Let $\alpha$ be sufficient for $(\iM,\iS)$ and let
$$
\iN_1=\{ a\in \iN\,:\, \alpha^*_{\omega}\circ\alpha(a)=a\}=\{a\in \iN_{\alpha}\,:\,
\alpha(\sig_t^{\omega\circ\alpha}(a))=\sig_t^\omega(\alpha(a))\}.$$
Then $\alpha(\iN_1)$
is a sufficient subalgebra and by Theorem \ref{T:4} and \ref{thm:3},
$\sigma_t^{\omega}(a(\theta))\in
\alpha(\iN_1)$ for all $t$, $\theta$, in particular, 
$a(\theta)=\alpha(b(\theta))$, for some elements  
$b(\theta)\in \iN_1$.  Consider the expansion:
\begin{eqnarray*}
[D\omega^{\alpha(b(\theta))},D\omega]_t
&=&\sum_{n=0}^\infty i^n\int_0^tdt_1\dots
\int_0^{t_{n-1}}dt_n\sig_{t_n}^\omega(\alpha(b(\theta)))...\sig_{t_1}^\omega
(\alpha(b(\theta)))\\
&=& \sum_{n=0}^\infty i^n\int_0^tdt_1\dots
\int_0^{t_{n-1}}dt_n\alpha(\sig_{t_n}^{\omega\circ\alpha}(b(\theta)))...\alpha(
\sig_{t_1}^{\omega\circ\alpha}(b(\theta)))\\
&=& \alpha([D\omega\circ\alpha^{b(\theta)},D\omega\circ\alpha]_t).
\end{eqnarray*}

On the other hand, $\alpha$ is sufficient, therefore
$[D\vfi_\theta,\omega]_t\in \alpha(\iN_\alpha)$ and
$$
\alpha([D\vfi_\theta\circ\alpha,D\omega\circ\alpha]_t)=
[D\vfi_\theta,D\omega]_t\, .
$$
As $\alpha$ is invertible on $\iN_\alpha$, it follows that
$[D\vfi_\theta\circ\alpha,D\omega\circ\alpha]_t=
[D[\omega\circ\alpha^{b(\theta)}],D\omega\circ\alpha]_t$
and we have (\ref{E:exsuff}).

Conversely,
suppose (\ref{E:exsuff}) holds, then
$$
\partial_jc(\omega\circ\alpha,b(\theta))=
-[\omega\circ\alpha^{b(\theta)}](\partial_jb(\theta))=
-\vfi_\theta(\alpha(\partial_jb(\theta))=
\partial_jc(\omega,\alpha(b(\theta)))
$$
for all $\theta$ and $j$. Putting $\theta=\theta_0$, it follows that
$c(\omega\circ\alpha,b(\theta))=c(\omega,\alpha(b(\theta)))$ for all $\theta$.
Hence
$$
S(\vfi_\theta||\omega)=c(\omega,\alpha(b(\theta)))-
\vfi_\theta(\alpha(b(\theta)))=c(\omega\circ\alpha,b(\theta))-
\vfi_\theta\circ\alpha
(b(\theta))=S(\vfi_\theta\circ\alpha||\omega\circ\alpha)
$$
and $\alpha$ is sufficient.
\qed

Note that the above Theorem implies, that if $\iS$ is the exponential family 
(\ref{E:qef}) for some $a_1,\dots, a_k\in \iM^{sa}$, then the coarse-graining 
is sufficient if and only if $\vfi_\theta\circ\alpha$ is again an 
exponential family, $\vfi_\theta\circ\alpha=[\omega\circ\alpha^{\sum_i
\theta_ib_i}]$ and $a_i=\alpha(b_i)$.  In  finite dimensions, the Theorem 
reduces to equality (\ref{E:matsuff}).

Let us denote
$$
\ell_i= \partial_i\left(a(\theta)-c(\omega,a(\theta))\right)=\partial_i
a(\theta)-\vfi_\theta(\partial_ia(\theta))
$$
Then $\ell_i$ is a quantum version of the {\bbff score} in classical 
statistics. We define  a Riemannian  metric tensor on $\Theta$ by
$$
g_{i,j}(\theta)=\gamma_{\vfi_\theta}(\ell_i,\ell_j)
$$
This is one of the quantum versions of the {\bbff Fisher information}, 
\cite{petz2002}. Note that $g_{i,j}(\theta)=\gamma_{\vfi_\theta}(\partial_ia(\theta),
\partial_ja(\theta))$ and  if $a(\theta)$ is twice differentiable, then
$$
g_{i,j}(\theta)=-\partial_i\partial_j
c(\omega,a(\theta))+
\vfi_\theta(\partial_i\partial_ja(\theta))
$$

Next we show how sufficiency can be  characterised  by the Fisher
information.

\begin{thm} 
Let $\alpha:\iN\to\iM$ and $\iS$ be as in the previous Theorem. 
Let $g(\theta)$ and $h(\theta)$  be the  Fisher information matrix 
for $\iS$ and the induced family $\{\vfi_\theta\circ\alpha\,:\,\theta
\in\Theta\}$, respectively. Then the matrix inequality
$$
h(\theta) \leq g(\theta)
$$
holds. Moreover, equality is attained if and only if $\alpha$ is sufficient 
for $(\iM,\iS)$.
\end{thm}

{\it Proof.} 
Let $c=(c_1,\dots,c_k)\in\mathbb R^k$, we have to show that 
$$
\sum_{i,j}c_ic_jh_{i,j}(\theta)\leq \sum_{i,j}c_ic_jg_{i,j}(\theta)
$$ 
for all $\theta$. Let $\vfi_\theta=[\omega^{a(\theta)}]$,  $\vfi_\theta
\circ\alpha=[\omega\circ\alpha^{b(\theta)}]$ and let us denote 
$$
\dot b=\frac d{dt}b(\theta+tc)|_{t=0}\in \iN,\quad   
\dot a=\frac d{dt}a(\theta+tc)|_{t=0}\in \iM.
$$
We have
\begin{eqnarray*}
\sum_{i,j}c_ic_jh_{i,j}(\theta)&=&
\gamma_{\vfi_\theta\circ\alpha}(\dot b,\dot b)=
-\frac d{dt}[\omega\circ\alpha^{b(\theta+tc)}](\dot b)\Big|_{t=0}
\cr &=&
-\frac d{dt}\vfi_{\theta+tc}\alpha((\dot b))\Big|_{t=0}=\gamma_{\vfi_\theta}
(\dot a, \alpha(\dot b))
\end{eqnarray*}
By Schwarz inequality and monotonicity of $\gamma$, we get
$$
\gamma_{\vfi_\theta}(\dot a, \alpha(\dot b))^2\leq \gamma_{\vfi_\theta}(\dot
a,\dot a)\gamma_{\vfi_\theta}(\alpha(\dot b),\alpha(\dot b))\leq 
\gamma_{\vfi_\theta}(\dot
a,\dot a)\gamma_{\vfi_\theta\circ\alpha}(\dot b,\dot b).
$$
This implies that
$$
\sum_{i,j}c_ic_jh_{i,j}(\theta)\leq \gamma_{\vfi_\theta}(\dot
a,\dot a)=\sum_{i,j}c_ic_jg_{i,j}(\theta).
$$

Suppose that $\alpha$ is sufficient, then there is a coarse-graining 
$\beta:\iM\to\iN$, such that $\vfi_\theta=\vfi_\theta\circ\alpha\circ\beta$ 
and, by the first part of the proof, $g(\theta)\leq h(\theta)$, 
hence $g(\theta)=h(\theta)$.

Conversely, let $g(\theta)=h(\theta)$, and let us denote
$a_i=\partial_ia(\theta)|_\theta$ and $b_i=\partial_ib(\theta)|_\theta$.
Then
$$
\partial_i\vfi_\theta(a_j)|_\theta=-g_{i,j}(\theta)=-h_{i,j}(\theta)=
\partial_i\vfi_\theta\circ\alpha(b_j).
$$
It follows that
$$
0=\partial_i\vfi_\theta(\alpha(b_j)-a_j)|_\theta=\gamma_{\vfi_\theta}(a_i, 
a_j-\alpha(b_j))
$$
for all $i$, $j$ and $\theta$. Therefore, we have for all $i$ and $\theta$,
$$
\gamma_{\vfi_\theta}(\alpha(b_i),\alpha(b_i))=\gamma_{\vfi_\theta}
(\alpha(b_i)-
a_i,\alpha(b_i)-a_i)+\gamma_{\vfi_\theta}(a_i,a_i).
$$
On the other hand, by monotonicity and the assumption, we have
$$
\gamma_{\vfi_\theta}(\alpha(b_i),\alpha(b_i))\leq
\gamma_{\vfi_\theta\circ\alpha}(b_i,b_i)=\gamma_{\vfi_\theta}(a_i,a_i)
$$
This implies that $\partial_i\alpha(b(\theta))-\partial_ia(\theta)=
\lambda_i(\theta)$ for some $\lambda_i(\theta)\in\mathbb R$, for all $i$ and
$\theta$. Since $a(\theta)$ and $b(\theta)$ are only determined up to a 
scalar multiple of $1$ and we may suppose that $b(\theta_0)=0$, 
$a(\theta_0)=0$,
we may choose $b(\theta)$ so that $a(\theta)=\alpha(b(\theta))$ for all
$\theta$. By Theorem \ref{thm:exsuff}, $\alpha$ is sufficient.
\qed

\section{Factorization}

Let  $\iM$ be a von Neumann  algebra with a standard representation on a Hilbert
space $\iH$  and let $\omega$ be a
faithful state on $\iM$. Let $\iM_0\subset \iM$ be a subalgebra and assume
that  it is invariant under the modular group $\sig_t^\omega$ of $\omega$.
Let $\omega_0$ be the restriction of $\omega$ to $\iM_0$, then
$\sigma_t^\omega|{\iM_0}=\sigma_t^{\omega_0}$.

Let $\phi$ and  $\phi_0$ be  faithful normal semifinite weights on $\iM$
and $\iM_0$, respectively, then for $a\in \iM_0$, we have
$$
\Delta_{\omega,\phi}^{it}a\Delta_{\omega,\phi}^{-it}=\sigma^\omega_t(a)=
\sig_t^{\omega_0}(a)=\Delta_{\omega_0,\phi_0}^{it}a
\Delta_{\omega_0,\phi_0}^{-it}
$$
It follows that there is a unitary  $w_t\in\iM_0'$, such that
$$
\Delta_{\omega,\phi}^{it}=\Delta_{\omega_0,\phi_0}^{it}w_t
$$

\begin{thm}\label{thm:fac}
Let $(\iM,\iS)$ be a statistical experiment dominated by a faithful normal 
state $\omega$. Let  $\iM_0\subset \iM$ be a von Neumann
subalgebra invariant with respect to the modular group $\sigma_t^\omega$. Then
$\iM_0$ is sufficient for $\iS$ if and only if for each $t\in\mathbb R$, there
is a unitary element $w_t\in\iM_0'$, such that
\begin{equation}\label{E:fac}
\Delta_{\vfi_\theta,\phi}^{it}=\Delta_{\vfi_{\theta,0},\phi_0}^{it}w_t,\quad
t\in\mathbb R
\end{equation}
where $\vfi_{\theta,0}=\vfi_\theta|_{\iM_0}$.
\end{thm}

{\it Proof.} Let $\iM_0$ be sufficient for $(\iM,\iS)$, then 
$[D\vfi_\theta,D\omega]_t=[D\vfi_{\theta,0},D\omega_0]_t$ for all $\theta$ and
$t$. It follows that
$$
\Delta_{\vfi_\theta,\phi}^{it}=[D\vfi_\theta,D\omega]_t
\Delta_{\omega,\phi}^{it}=
[D\vfi_{\theta,0},D\omega_0]_t\Delta_{\omega_0,\phi_0}^{it}w_t=
\Delta_{\vfi_{\theta,0},\phi_0}^{it}w_t
$$

Conversely, suppose (\ref{E:fac}), then
$$
[D\vfi_\theta,D\omega]_t=
\Delta_{\vfi_\theta,\phi}^{it}\Delta_{\omega,\phi}^{-it}=
\Delta_{\vfi_{\theta,0},\phi_0}^{it}w_tw_t^*\Delta_{\omega_0,\phi_0}^{-it}=
[D\vfi_{\theta,0},D\omega_0]_t
$$
and $\iM_0$ is sufficient.
\qed

Let $\iM_1=\iM_0'\cap \iM$ be the relative commutant, then
$\iM_1$  is invariant under $\sig_t^\omega$ as well and $\sig_t^\omega|{\iM_1}
=\sig_t^{\omega_1}$, where  $\omega_1=\omega|_{\iM_1}$. Suppose further, that
the subalgebra $\iM_0$ is semifinite and let $\phi_0$ be a trace. Then 
$\Delta^{it}_{\omega_0,\phi_0},\Delta_{\vfi_{\theta,0},\phi_0}^{it}\in\iM_0$
for all $\theta$ and there is an operator 
$\Delta$ affiliated with $\iM_0'$, such that $w_t=\Delta^{it}$. Moreover,
for $a\in\iM_1$,
$$
\sig^{\omega_1}_t(a)=\sig^\omega_t(a)=w_taw_t^*=\Delta^{it}a\Delta^{-it}
$$
The factorization (\ref{E:fac}) has a special form, if we require that 
the entropy of the state $\omega$ is finite.
Recall that the {\bbff entropy} of a state $\vfi$ of a C*-algebra is defined 
as
$$
S(\vfi):=\sup \Big\{ \sum_i \lambda_i S(\vfi_i\|\vfi):\sum_i\lambda_i \vfi_i=
\vfi\Big\},
$$
see (6.9) in \cite{OP}. If $S(\omega)<\infty$, then 
$\iM$
must be a countable direct sum of type I factors, see Theorem 6.10. in 
\cite{OP}. As the subalgebras $\iM_0$ and $\iM_1$ are invariant under $\sigma_t^\omega$, 
we have by Proposition 6.7. in \cite{OP}
that $S(\omega_0), S(\omega_1)\leq S(\omega)<\infty$. It follows that both 
$\iM_0$ and $\iM_1$ must be
countable direct sums of type I factors as well.

Let $\phi$ and $\phi_0$ be the canonical traces and let $\rho_\omega$, $\rho_\theta$ 
and $\rho_{\theta,0}$, $\rho_{\omega_0}$ be the density operators. 
Then $w_t=\rho_{\omega_0}^{-it}\rho_\omega^{it}\in \iM_0'\cap\iM=\iM_1$ and 
since $\sig^{\omega_1}_t(a)=w_taw_t^*$, we have $w_t=\rho_{\omega_1}^{it}z^{it}$
for a central element $z$ in $\iM_1$ and a density operator $\rho_{\omega_1}$ in
$\iM_1$. Putting all together, we get that sufficiency is equivalent with
\begin{equation}\label{E:fac2}
\rho_\theta=\rho_{\theta,0}\rho_{\omega_1}z,\qquad \theta\in\Theta
\end{equation}
The essence of this factorization is that the first factor
is the reduced density and the rest is independent of $\theta$.

Since $\iM_1$ is a countable direct sum
of factors of type I,  there is an orthogonal
family of minimal central projections $p_n$, $\sum_np_n=1$.
Moreover, there is a decomposition
$$
\iH_n=p_n\iH=\iH^L_n\otimes\iH_n^R,
$$
such that
$$
\iM_1 =\bigoplus_n \bbbc I_{\iH^L_n}\otimes B(\iH^R_n),\qquad
(\iM_1)'=\bigoplus_nB(\iH^L_n)\otimes \bbbc I_{\iH^R_n}
$$
{F}rom  this, we get
\begin{equation}\label{eq:factortheta2}
\rho_\theta=\rho_{\theta,0}\rho_{\omega_1}z=\sum_n \vfi_\theta(p_n)\rho^L_n(\theta)\otimes 
\rho^R_n\,
\end{equation}
where $\rho^R_n$ is a density operator in $B(\iH^R_n)$ and $\rho^L_n(\theta)$ is a
density operator in $B(\iH^L_n)$. 

A particular example of a sufficient subalgebra is the subalgebra generated
by the partial isometries $\{[D\vfi_\theta,D\omega]_t:
t\in \bbbr\}$, this subalgebra is  minimal sufficient and
invariant under $\sig_t^\omega$. If $S(\omega)<\infty$, the decomposition
(\ref{E:fac2}), corresponding to this subalgebra is a maximal such
decomposition, in the sense that the density operator  $\rho_{\theta,0}$ cannot be
decomposed further, in a nontrivial way.

\begin{ex}  Let $\iH$ be a finite dimensional Hilbert space, let
$\mathcal S$ be a family of pure states induced by the unit vectors
$\{\xi_\theta\,:\,\theta\in \Theta\}$. Suppose that the vectors 
$\xi_\theta$ generate $\iH$, then there is a faithful
state $\omega$, dominating $\iS$. Let 
$$
\mathcal A_0=\oplus_{j=1}^mB(\iH_j^L)\otimes \bbbc I_{\iH_j^R}
$$ 
be a subalgebra in $B(\iH)$, invariant under $\sigma^\omega_t$
and suppose that $\mathcal A_0$ is sufficient for $\iS$.
Then, we have from (\ref{eq:factortheta2}) that for each $\theta$, 
there is some $1\leq j\leq m$ and unit vectors $\xi_{\theta,j}\in \iH^L_j$, 
$\xi_j\in \iH^R_j$, such that
$$
\xi_\theta=\xi_{\theta,j}\otimes\xi_j
$$
Suppose that there are $\theta_1$, $\theta_2 \in \Theta$, such that
$\xi_{\theta_i}=\xi_{\theta_i,j_i}\otimes \xi_{j_i}$, $i=1,2$ and
$j_1\neq j_2$, then $\xi_{\theta_1}$ and $\xi_{\theta_2}$ must be
orthogonal. Consequently, if, for example, the family $\mathcal S$ contains
no two orthogonal vectors, then $m=1$, $\mathcal A_0$ must be of the form
$\mathcal A_0=B(\iH_L)\otimes \bbbc I_{\iH_R}$
and $\xi_\theta=\xi_{\theta,L}\otimes\xi_R$ for all $\theta$. \qed
\end{ex}

\begin{ex} 
Let us return to the experiment $(\iM^{\otimes n},\{\vfi^{\otimes
n}_\theta\})$ of Example \ref{ex:tensor}. Let $\iM=B(\iH)$ and let $\pi$
be the unitary representation of the permutation group $S(n)$ on 
$\iH^{\otimes n}$, then $\iN=\pi(S(n))'$. There is a decomposition 
$\pi=\oplus_{i,j}\pi_{i,j}$, such that all $\pi_{i,j}$ are irreducible
representations and  $\pi_{i,j}$, $\pi_{k,l}$ are equivalent if and only if
$i=k$. It follows that there is a decomposition $\iH^{\otimes n}= \oplus_k
\iH_k^L\otimes \iH_k^R$ such that
$$
\iN=\bigoplus_k B(\iH_k^L)\otimes \bbbc I_{\iH_k^R}
$$
Let $\omega$ be a state dominating $\vfi_\theta$, $\theta\in \Theta$, then 
$\omega^{\otimes n}$ dominates $\vfi^{\otimes n}$, $\theta\in\Theta$.
Since $\iN$ is also invariant under the modular group $\sigma^{\omega^{\otimes
n}}_t$, we conclude that the densities decompose as
$$
\rho_\theta^{\otimes n}=\sum_k \lambda_k \rho^L_k(\theta)\otimes \rho^R_k,
$$
for density matrices $\rho^L_k(\theta)\in B(\iH_k^L)$ and $\rho^R_k\in B(\iH^R_k)$. \qed
\end{ex}

\end{document}